\documentstyle[12pt,epsfig]{article}

\newcommand{\beq}{\begin{equation}}
\newcommand{\eeq}{\end{equation}}
\newcommand{\bea}{\begin{eqnarray}}
\newcommand{\eea}{\end{eqnarray}}
\def\half{{\textstyle{1\over 2}}}

\def\as{\alpha_S}
\def\df{\Delta\phi}

\def\pti{p_{T1}}
\def\ptii{p_{T2}}
\def\vpti{\vec{p}_{T1}}
\def\vptii{\vec{p}_{T2}}
\def\ptisq{p_{T1}^2}
\def\ptiisq{p_{T2}^2}

\def\d{{\rm d}}

\def\rsone{\sqrt s_1}
\def\rstwo{\sqrt s_2}

\def\Dzero{D$\emptyset$}

\def\TeV{{\rm TeV}}
\def\GeV{{\rm GeV}}

\def\cR{{\cal R}}

\def\lapprox{\lower .7ex\hbox{$\;\stackrel{\textstyle <}{\sim}\;$}}
\def\gapprox{\lower .7ex\hbox{$\;\stackrel{\textstyle >}{\sim}\;$}}

\begin{document}
\titlepage
\begin{flushright}
{DTP/98/40}\\
{UR-1532}\\
{hep-ph/9806371}\\
{June 1998}\\
\end{flushright}
\begin{center}
\vspace*{2cm}
{\Large {\bf BFKL Physics in Dijet Production at the LHC }} \\

\vspace*{1.5cm}
Lynne H.\ Orr$^a$ and W.\ J.\ Stirling$^{b}$ \\

\vspace*{0.5cm}
$^a \; $ {\it Department of Physics and Astronomy, University of Rochester,
Rochester, NY~14627-0171, USA}\\

$^b \; $ {\it Departments of Physics and  Mathematical Sciences, 
University of Durham,
Durham, DH1~3LE, UK}\\

\end{center}

\vspace*{4cm}
\begin{abstract}
The production in hadron-hadron collisions of jet pairs with
large rapidity separation and comparable modest
transverse momentum is, in principle, described by the  perturbative QCD
BFKL equation.
The measurement of such jet pairs appears
well suited to the LHC detectors, with their
ability to detect forward jets. We present predictions for dijet 
cross sections and
correlations obtained using a BFKL Monte Carlo which allows
kinematic and other subleading effects to be incorporated.
The enhanced phase space for gluon emission at the LHC makes the
BFKL behavior somewhat easier to observe than at the Tevatron,
although kinematic effects are still important.
The production of forward jets in association with heavy Higgs
bosons via the gauge boson fusion mechanism is also studied
and compared with QCD dijet production.
\end{abstract}

\newpage

\section{Introduction}

In perturbative QCD physical quantities are typically computed in fixed-order
expansions in powers of the coupling constant $\as$.  In some
kinematic regimes, however, large logarithms multiply the coupling 
and the series must be resummed to all orders.  In certain cases
this resummation is possible and meaningful predictions can still
be obtained.  For example, large logarithms in the  small-$x$ region in 
deep inelastic lepton-hadron scattering are resummed by the
Balitsky, Fadin, Kuraev and Lipatov (BFKL)  equation \cite{bfkl},
to yield the well-known BFKL prediction $F_2 \sim (1/x)^\lambda$,
with  $\lambda = 4C_A\ln 2\, \as/\pi \approx 0.5$.

\begin{figure}[t]
\begin{center}
\mbox{\epsfig{figure=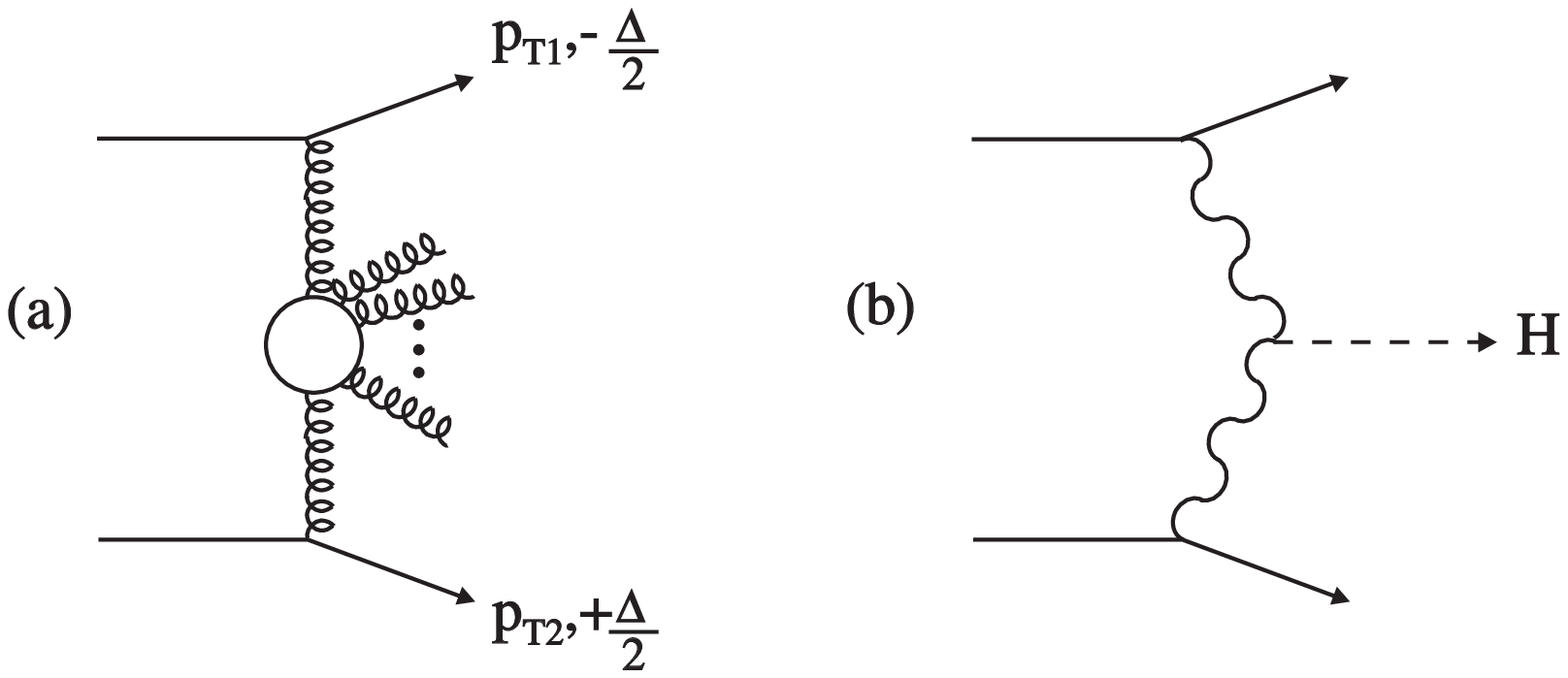,width=16.0cm}}
\caption[]{Schematic representation of dijet production at large
rapidity separation $\Delta$ in hadron-hadron collisions in (a) the
BFKL approach and (b) Higgs production via boson fusion.}
\label{fig:diagrams}
\end{center}
\end{figure}

In hadron collisions BFKL resummation also applies to production of jet pairs
at large rapidity separation $\Delta$ when the jets have comparable transverse
momenta that are small compared to the center-of-mass energy.  At large values
of $\Delta$, the cross section receives contributions from  
real and virtual gluons  emitted in the rapidity interval between the two 
jets,
as illustrated in Fig.~\ref{fig:diagrams}(a).  The BFKL equation resums the
leading logarithmic contributions from these gluons \cite{muenav}, giving
rise to a  cross section that increases with $\Delta$, 
$\hat\sigma\sim\exp(\lambda \Delta)$, with $\lambda$ defined above.  
Mueller and
Navelet \cite{muenav} pointed out the potential of this increase, which
does not appear in lowest-order QCD, to provide a way to
investigate BFKL physics at hadron colliders by looking at dijet production
at large rapidity separation.

Unfortunately the $\Delta$ increase disappears when the subprocess 
cross section is folded in with parton distribution functions (pdfs), which
decrease with $\Delta$ more rapidly than $\hat\sigma$ increases.  
Several methods have been suggested to avoid this pdf sensitivity 
when studying  `Mueller-Navelet'  jets.  One \cite{ds,wjs,os}
is to study the 
decorrelation in the two jets' azimuthal angles  that arises from
emission of gluons between the jets; BFKL predicts a stronger decorrelation
than does fixed-order QCD, and this prediction should be relatively 
insensitive to the pdfs.  Another \cite{muenav,osbis} is to look for the
$\Delta$ increase in the ratio of dijet cross sections at two 
center-of-mass energies for corresponding parton momentum fractions, so
that the pdfs cancel. 

In practice it is not useful to compare
analytic asymptotic BFKL predictions  directly
with experiment because nonleading corrections can be large.  In particular,
in the analytic BFKL calculation
gluons can be emitted arbitrarily, with no kinematic
cost, and energy and momentum are not conserved.  In Refs.~\cite{os,osbis}
(see also \cite{schmidt}) a Monte Carlo 
approach is used 
in which the emitted gluons are subject to kinematic 
constraints, and other
nonleading effects such as the running of $\as$ are included.
Kinematic constraints are seen to have a significant effect, suppressing
the emission of large numbers of energetic gluons.  The studies in 
\cite{os} and \cite{osbis} focused on the azimuthal decorrelation and
cross section ratios at the Fermilab Tevatron $p\bar p$ collider.

Here we are concerned with dijet production at the CERN LHC $pp$ collider.
The phase space available for studying dijet production is larger at the
LHC than at the Tevatron, because the  center-of-mass energy is much
higher while measurable jet transverse momentum thresholds increase only
slightly, if at all.  As a result we expect kinematic constraints to 
have  weaker effects  at the LHC, so that the predictions of 
naive BFKL should be more robust than at the Tevatron.  We
will address this in more detail below.

BFKL physics (or, more generically, QCD) is not the only potential source of 
jet pairs with large rapidity separation at the LHC:  heavy Higgs production 
via gauge boson fusion in the process $qq\to qqH$
(see Fig.~\ref{fig:diagrams}(b))
is well known to give rise to forward jets \cite{wwfus,cahn}.
In fact the forward calorimetry in the LHC detectors has been designed to 
be able to detect jets at large rapidity  
for the purpose of observing this process \cite{lhcdet}.
Detection of jets with rapidity $\vert y \vert \lapprox 5$
(i.e. $\Delta \lapprox 10$) and transverse momentum
$p_{T} > {\cal O}(20\; \GeV)$ should be possible.

 In contrast to the BFKL
case, where the region between the jets is populated by gluons, the 
jets produced in association with the Higgs are characterized by an
absence of hadronic activity between them --- a `rapidity gap.' 
We will see other
differences when we compare other features of dijet production in the two 
cases below. The Higgs process therefore acts as a useful comparator 
for the BFKL study. 

The paper is organized as follows.  In the next section we summarize the 
essential features of the BFKL MC calculation used to obtain our results.  
In section 3 we present results for dijet production in $pp$ collisions
at center-of-mass energy 14 TeV, first in the BFKL approach, and 
then for jets produced in association with a  heavy Higgs boson.
In section 4 we investigate the collision energy dependence of the BFKL 
dijet cross section and present the ratio of cross sections at 10 and 
14 TeV.  In section 5 we present our conclusions.

\section{Improved BFKL approach to dijet production in had\-ron collisions}

The differential subprocess cross section for production of a jet pair 
with rapidity difference $\Delta\equiv \vert y_1-y_2\vert $ and relatively 
small transverse momenta $p_{T1},p_{T2}$ is given in the BFKL approach
by 
\beq
{d \hat{\sigma} \over d^2 p_{T1}d^2 p_{T2} d \Delta}
 = {\as^2  C_A^2 \over \ptisq \ptiisq }\;
   f(\vpti, \vptii, \Delta).
\label{eq:sighat}
\eeq
The Laplace transform  of the function $f$ with respect to $\Delta$
satisfies the BFKL equation.   This equation contains integrals over
both real and virtual gluons, and when its Laplace-transformed  solution
is substituted into Eq.~(\ref{eq:sighat}), it gives the sum over emitted
gluons and total cross section behavior discussed above.  This is the 
`naive' BFKL case.

As indicated in the introduction, the price of having an analytic solution is 
integration over radiated gluons with arbitrarily large transverse energies.
We can improve on this approach by unfolding the implicit sum
over gluons that appears in the analytic solution.  
Details can be found in \cite{os}; see also \cite{schmidt} for a similar 
approach.
The basic idea is that 
real gluons are separated into `resolved' and `unresolved' gluons,
the latter being those that fall below a transverse momentum  threshold
$\mu_0$ and
are undetectable as jets.  The unresolved gluon contributions are then combined 
with those from virtual contributions to give an overall form factor that
suppresses the cross section.  The cross section is then an explicit sum over
resolved real gluons; for fixed $\as$ it becomes
\beq
{d \hat{\sigma}_{gg} \over d^2 p_{T1}d^2 p_{T2} d \Delta}
 = {\as^2  C_A^2 \over \ptisq \ptiisq }\;
   \left[ {\mu_0^2\over \ptisq}\right]^{\as C_A\Delta/\pi}\;
    \sum_{n=0}^{\infty} \cR^{(n)}(\vpti, \vptii, \Delta )\;  ,
\label{eq:sighatos}
\eeq
with $\cR^{(0)} = \delta( \vpti +  \vptii )$ and $\cR^{(n)}$ for $n>0$ given
in \cite{os}, where the expressions for running $\as$ can also be found.
The analytic result for the total cross section is reproduced in
the limit $\mu_0\to 0$.\footnote{In practice, the value $\mu_0 = 1$~GeV
gives a very good approximation to the limiting result and acceptable
Monte Carlo efficiency. In general $\mu_0$ should be small compared to 
$\pti, \ptii$.}

It is straightforward to implement this
in a Monte Carlo 
program where kinematic cuts can be applied directly.  The total cross
section is obtained by folding in the subprocess cross section 
(\ref{eq:sighatos}) with the appropriate parton distributions
functions; see \cite{os} for details.

\section{Jet pair production in $pp$ collisions
at $\protect\sqrt{s}=14\; \protect\TeV$}

\subsection{BFKL calculation of dijet production}

The results shown here were obtained using the BFKL Monte Carlo program 
described above and in Ref.~\cite{os}.  We consider dijet production in
which the jets have equal and opposite rapidity $y_1=-y_2=\Delta/2$
and transverse momentum greater than some value $P_T$.\footnote{Equal
and opposite rapidities are taken  for convenience only and are
not required for validity of the BFKL approach; the rapidity
separation is the important quantity.}
For the LHC we
consider transverse momentum thresholds $P_T=20\; \GeV$ and $P_T=50\; \GeV$.
Given the high level
of hadronic activity at the LHC, the former value may be slightly
optimistic but the latter is certainly achievable \cite{lhcdet}.  We use CTEQ4L
parton distributions \cite{cteq4}.

\begin{figure}[t]
\begin{center}
\mbox{\epsfig{figure=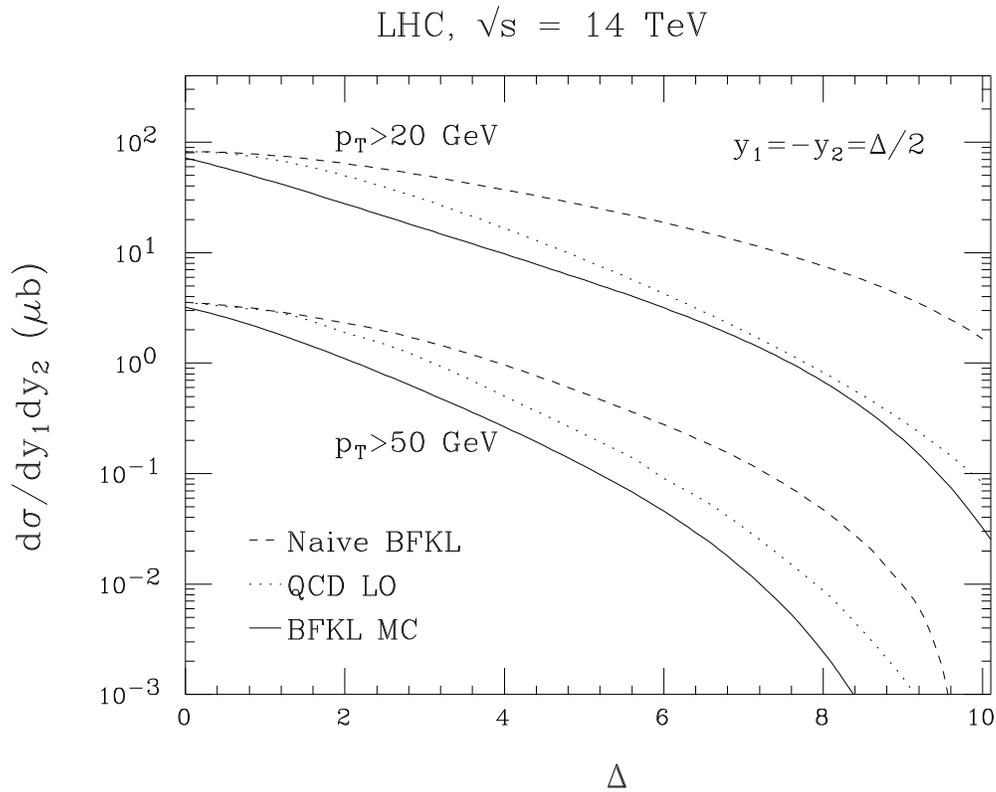,width=16.0cm}}
\caption[]{
BFKL and asymptotic QCD leading-order 
dijet production cross sections at the LHC ($\sqrt{s}=14\; \TeV$)
as a function of 
the dijet rapidity separation. The pdfs are the CTEQ4L set
 \protect\cite{cteq4}. The three curves at each transverse momentum 
 threshold 
 use: (i) improved BFKL MC with running $\as$ (solid lines), 
 (ii) naive BFKL (dashed lines),
  and (iii) the asymptotic ($\Delta \gg 1$) form
 of QCD leading order (dotted lines).
}
\label{fig:bfklsig}
\end{center}
\end{figure}

Figure~\ref{fig:bfklsig} shows the dijet production  cross section
at the LHC with center-of-mass energy $14\; \TeV$ for $p_T>20\; \GeV$ (upper
curves) and 
$p_T>50\; \GeV$ (lower curves).  All of the curves fall off with increasing
$\Delta$ as a result of pdf suppression; note that the difference in momentum 
thresholds causes the $p_T>50\; \GeV$
cross sections to reach  the kinematic 
limit at smaller values of $\Delta$ than  the $p_T>20\; \GeV$ curves.

The naive BFKL prediction, obtained
using the analytic (asymptotic) BFKL solution combined with pdfs assuming 
lowest-order
kinematics, is shown as dashed curves in the figure.  For both $p_T$
thresholds, the naive BFKL cross section is largest for all values of $\Delta$;
this reflects the rise with $\Delta$ of the subprocess cross section.
The falloff with $\Delta$ is due to the pdfs.  The BFKL MC results, shown as
solid curves, lie well below naive BFKL as a result of the kinematic 
suppression.\footnote{The running of $\as$ also contributes to the
suppression but kinematics is the main effect.}  Clearly nonleading
effects due to kinematics are important.  Also shown for comparison are
the cross sections in the asymptotic (large $\Delta$) lowest-order QCD limit
(see \cite{osbis} for details about this limit).  It lies below the
naive BFKL curves, because the subprocess cross section is asymptotically
flat in $\Delta$, and it lies above the BFKL MC curves because it reflects only
leading-order kinematics, whereas the BFKL MC cross section pays the pdf penalty
for emitted gluons.  

A characteristic of BFKL dijets   that clearly distinguishes them from 
those in leading-order QCD is a decorrelation in azimuthal angle at
large values of $\Delta$ \cite{ds,wjs,os}.  In QCD at lowest order
the two jets are strictly back-to-back in the transverse plane,
so that their azimuthal angles are completely correlated,  independent
of rapidity.  In contrast,
in the BFKL case gluons are radiated in the rapidity interval between the
two jets, and as this interval increases more gluons can be radiated.
The presence of these gluons weakens the azimuthal correlation between  
the two jets of
interest, so that one sees an increasing azimuthal decorrelation as  
$\Delta$ increases.  

As discussed in \cite{os}, applying kinematic constraints as in the 
improved BFKL MC gives less of 
a decorrelation than predicted in naive BFKL because emission of arbitrarily 
many gluons of arbitrary energy is no longer allowed.  In fact, for any given
transverse momentum threshold, there is some $\Delta$ at which the jet
pair alone saturates the kinematic limit, and emission of additional 
(real) gluons is completely suppressed and the azimuthal correlation
returns.  As we approach that limiting value of $\Delta$
we therefore expect to see a transition back towards correlated jets.

\begin{figure}[t]
\begin{center}
\mbox{\epsfig{figure=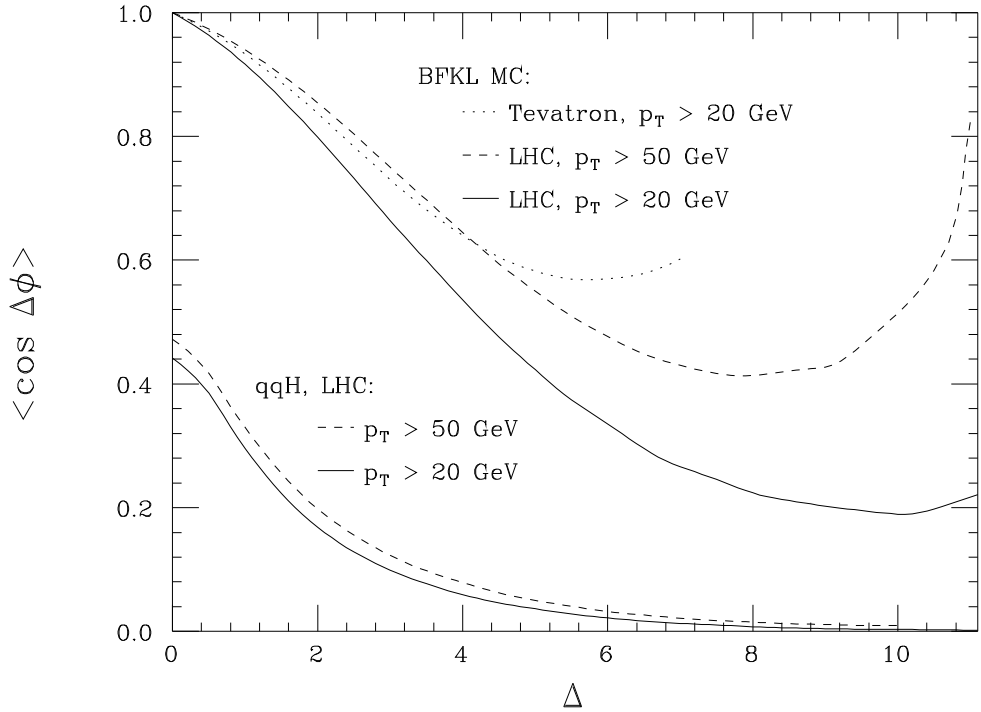,width=16.0cm}}
\caption[]{
The azimuthal angle decorrelation in dijet production at the Tevatron 
($\sqrt{s}=1.8\; \GeV$) and LHC ($\sqrt{s}=14\; \TeV$)
as a function of dijet rapidity difference $\Delta$.  
The upper curves are computed using the improved BFKL MC with running $\as$;
they are: (i) Tevatron, $p_T>20\; \GeV$ (dotted curve),
(ii) LHC, $p_T>20\; \GeV$ (solid curve), and (iii) LHC, $p_T>50\; \GeV$
(dashed curve).  The lower curves are for dijet production in the process
$qq\to qqH$ for $p_T>20\; \GeV$ (solid curve) and $p_T>50\; \GeV$
(dashed curve).
}
\label{fig:azidecorr}
\end{center}
\end{figure}

This is illustrated in Fig.~\ref{fig:azidecorr}, where we show 
the azimuthal decorrelation in dijet production in the improved BFKL
MC approach (upper curves).  We show the mean value of $\cos{\Delta\phi}$ 
where we have defined
\beq
\df\equiv \vert \phi_1-\phi_2 \vert -\pi
\eeq
so that $\df=0$ when the two jets are back-to-back in the transverse plane.
The jets are completely correlated at $\Delta=0$, and as $\Delta$
increases
we see the characteristic BFKL decorrelation, 
followed by a flattening out and then an increase in 
$\langle\cos{\Delta\phi}\rangle$
as we approach the kinematic limit.  Not surprisingly, the kinematic
constraints have a much stronger effect when the $p_T$ threshold is 
set at $50\; \GeV$ (dashed curve) than at $20\; \GeV$ (solid curve); 
in the latter case
more phase space is available to radiate gluons.  We also show for 
comparison the decorrelation for dijet production at the Tevatron
for $p_T>20\; \GeV$.  There we see that the lower collision energy (1.8~TeV)
limits the allowed rapidity difference and 
substantially suppresses the decorrelation at large $\Delta$.  The larger
center-of-mass energy compared to  transverse momentum threshold 
at the LHC would seem to give it a significant advantage as far 
as observing BFKL effects is concerned.

The lower set of curves in Fig.~\ref{fig:azidecorr} refer to Higgs production
and will be discussed below.

\subsection{Forward jets from Higgs production via boson fusion}

At the LHC, Higgs bosons can be produced in the process $qq\to qqH$,
where $W$ or $Z$ bosons emitted by the initial state quarks annihilate
to form a Higgs \cite{wwfus}, see Fig.~1(b). This mechanism may be important
for discovering a heavy $M_H \sim 1$~TeV Higgs, since the rate there
is comparable to the rate from the `standard' $gg\to H$ process, see for
example Ref.~\cite{qcdbook}. For lighter
Higgs bosons, weak boson fusion can provide a complementary method of
checking that the couplings to heavy quarks and weak bosons are related
in the way specified by the Standard Model. 

Our interest here, however,
 is in the final-state quark jets accompanying the Higgs. The ability 
 to `tag' these jets \cite{cahn}  is necessary to suppress backgrounds,
 for example from $q\bar q\to WW,\ ZZ,\ \mbox{and}\ t \bar t$ production.
It is the $t$-channel $W$ and $Z$ propagators (see Fig.~1(b))
that  to a large extent determine
the regions of phase space populated by the tag  jets. Typically, they 
have ${\cal O}(\TeV)$ energies and transverse momenta
less than $M_W$. They therefore naturally 
have large (roughly equal and opposite) rapidities.

\begin{figure}[t]
\begin{center}
\mbox{\epsfig{figure=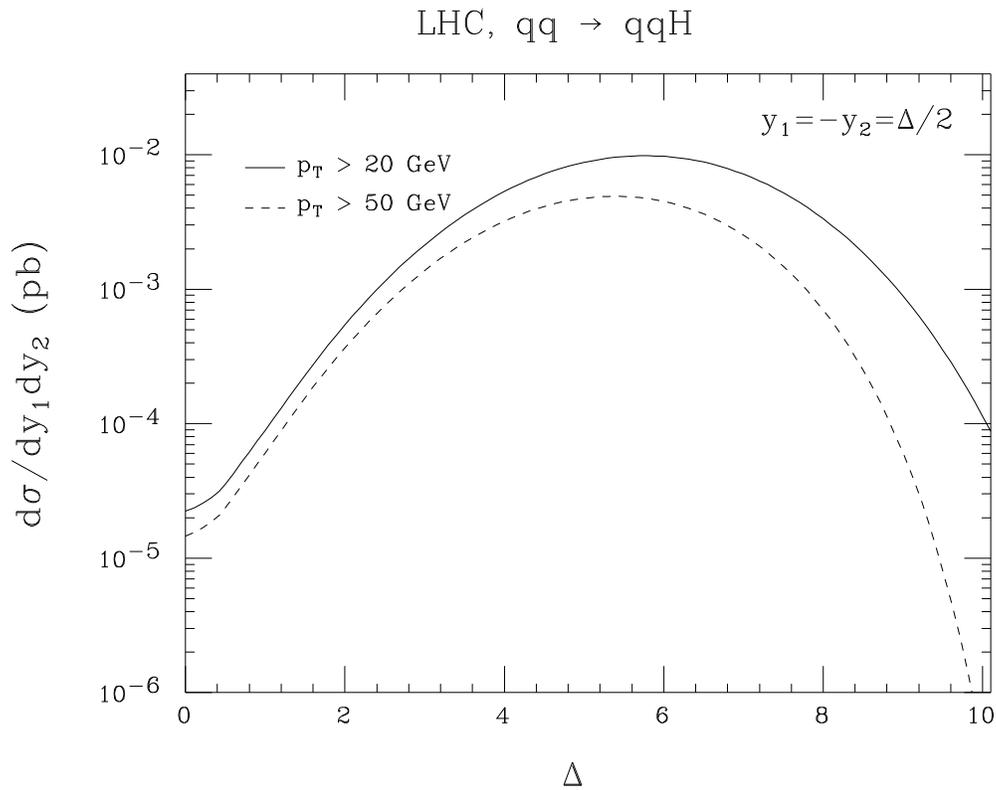,width=16.0cm}}
\caption[]{The cross section for the production of  jet pairs in association
with a Higgs boson via gauge boson fusion (Fig.~\ref{fig:diagrams}(b)) at the 
LHC ($\sqrt{s}=14\; \TeV$) as
a function of rapidity separation of the jet pair for $p_T>20\; \GeV$
(solid curve), and  $p_T>50\; \GeV$ (dashed curve).  The pdfs are the CTEQ4L set
 \protect\cite{cteq4}.
}
\label{fig:qqhsig}
\end{center}
\end{figure}

Higgs production via $WW$ and $ZZ$ fusion therefore automatically provides
a  `BFKL-like' dijet sample with large rapidity separation. 
Figure~\ref{fig:qqhsig} shows the 
differential cross section $d^2\sigma^H/dy_1dy_2(y_1 = -y_2 = \Delta/2)$
as a function of $\Delta$, for $M_H = 500$~GeV and
$p_{T} > 20,\ 50$~GeV at the LHC. Notice the difference
in shape compared to the QCD LO dijet distributions shown in 
Fig.~\ref{fig:bfklsig}; 
the tag jets are naturally produced with a sizeable rapidity separation.
The big difference between  Higgs and QCD dijet production is 
of course the $t$-channel color {\it singlet} exchange property of the former.
This gives rise to a rapidity {\it gap}, rather than a region between the jets
populated by BFKL soft gluons, Fig.~1(a).\footnote{The existence of this
 rapidity gap has been suggested as an additional signature for Higgs
 production, see for example Ref.~\cite{higgsgap}.} The 
 large rapidity separation decorrelation between 
 the forward jets induced by multigluon emission between them is therefore
 absent. Interestingly, however, for $M_H \gg p_{T}(\mbox{jet})$
the centrally produced
 Higgs boson acts as a jet decorrelator, i.e. the $qq\to Hqq$ matrix element
 does not depend on the relative orientation of $\vec{p_{T1}}$ and 
$\vec{p_{T2}}$. Indeed the only significant correlation occurs when the
tag jets are produced centrally (small $\Delta$) and all invariants 
$p_i\cdot p_j$ are large and of the same order. Three-body kinematics
then yields approximately back-to-back jets. The situation is summarised
in Fig.~\ref{fig:azidecorr}, which shows the $\Delta$ dependence  of
$\langle\cos\Delta\phi\rangle$
for the $qq\to qqH$ process (lower set of curves). 
We have again taken $M_H = 500$~GeV, but in  
fact the dependence on the (heavy) Higgs mass is very weak. Evidently
the jets accompanying Higgs production are much {\it more} decorrelated in 
azimuth than the Mueller-Navelet QCD jets. In principle, therefore,
the Higgs tag jets could act as a benchmark for the decorrelation,
since the theoretical predictions shown in Fig.~\ref{fig:azidecorr} 
should be reliable.

\section{Collision energy dependence of dijet production at the LHC}

The BFKL increase in the dijet subprocess cross section with $\Delta$
is swamped by the decreasing parton distribution functions, as can
be seen in Fig.~\ref{fig:bfklsig}.  One way \cite{muenav,osbis}
to account for this
is to take a ratio of cross sections at different $\Delta$ but at the same
values of parton momentum fraction $x$ so that the pdf dependence
cancels out.  This requires (at least) two
center-of-mass energies.  The difficulty with this method,  as
discussed in \cite{osbis}, is that the cancellation of pdfs only occurs
using lowest-order kinematics; like any other quantity, the cross
section ratio is subject to corrections due to kinematic effects.
As shown in \cite{osbis}, these kinematic effects turn out to be 
very important
at the Tevatron (for $\sqrt{s}=630$ and $1800\; \GeV$)
and the improved BFKL MC prediction differs qualitatively from 
that of naive BFKL.  

In this section we investigate the dijet cross section ratio at the LHC, 
for the design center-of-mass energy $\sqrt{s}=14\; \TeV$ and for
$\sqrt{s}=10\; \TeV$, which has been mentioned as a possible first step
toward the design energy.
At the LHC,  the higher center-of-mass energy
compared to transverse momentum threshold may mitigate the strong kinematic
suppression that spoiled the naive BFKL ratio at the Tevatron.

We consider a cross section ratio $R_{12}$ at center-of-mass energies
$\sqrt{s_1}$ and $\sqrt{s_2}$ measured at fixed values of rapidity difference
$\Delta_1$ and $\Delta_2$ chosen so that the asymptotic lowest-order
QCD cross sections (including pdfs) are equal \cite{osbis}.  Given $\Delta_1$,
$\Delta_2$ is given by
\beq
{ \cosh(\half\Delta_1)  \over  \cosh(\half\Delta_2) }
= { \rsone \over \rstwo }. 
\label{deltadef}
\eeq
The cross section ratio to be measured is then 
\beq
R_{12} ={  \d\sigma(\rsone,\Delta_1) \over  \d\sigma(\rstwo,\Delta_2)}
\label{ratiodef}
\eeq
One advantage of this definition is that it uses the rapidity
differences between the two outside jets, rather than the parton momentum
fractions for the event, which can be more difficult to
measure \cite{ANNA}.

\begin{figure}[t]
\begin{center}
\mbox{\epsfig{figure=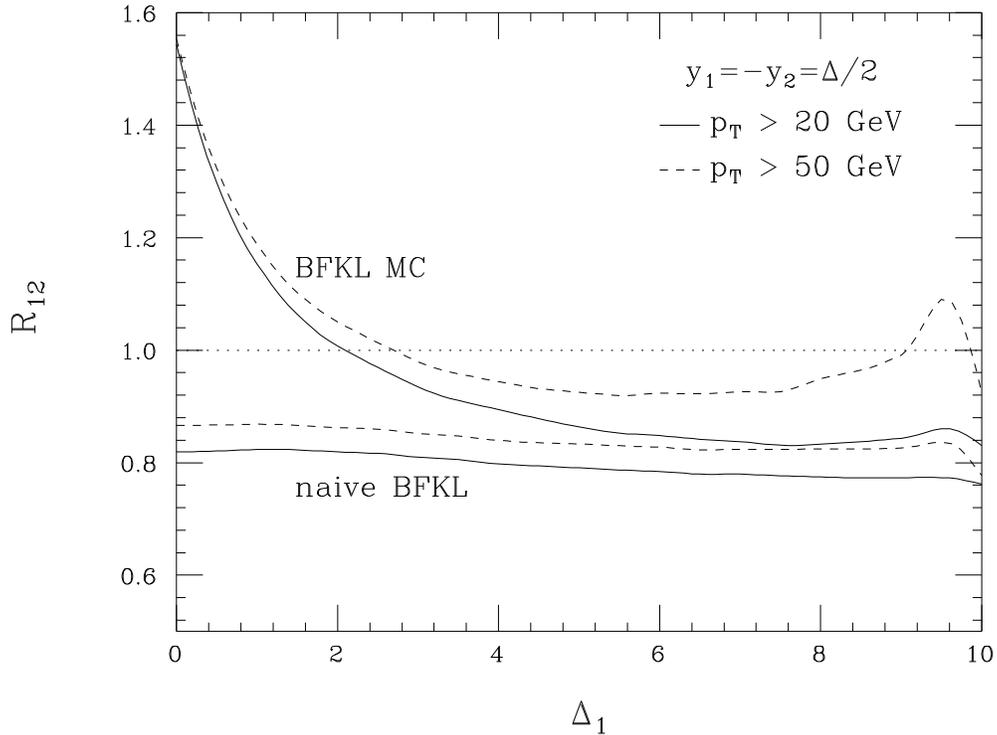,width=16.0cm}}
\caption[]{
The  ratio $R_{12}$ of the dijet cross sections at the two collider energies
$\rsone = 10\; \TeV$ and $\rstwo=14 \; \TeV$, as defined in the text
for $p_T>20\; \GeV$
(solid curves), and  $p_T>50\; \GeV$ (dashed curves).
The curves are calculated according to: (i) the improved BFKL MC 
predictions using
CTEQ4L \cite{cteq4}  pdfs (upper curves), 
 with $\mu =P_T = 20\; \GeV$,  
(ii) the naive BFKL prediction (lower curves), and 
(iii)  the
asymptotic leading-order prediction (dotted curve)  $R_{12}=1$.
}
\label{fig:ratio}
\end{center}
\end{figure}

In Fig.~\ref{fig:ratio} we show $R_{12}$ at the LHC with $\rsone=10\; \TeV$
and $\rsone=14\; \TeV$ as a function of $\Delta_1$, with $\Delta_2$ determined
by Eq.~(\ref{deltadef}).  Note that although we show results for all values of
$\Delta$,
it is really only at the larger values that we expect BFKL behavior to
be manifest.
In  asymptotic LO QCD $R_{12}=1$  by construction; it is shown for reference
as a dotted line in the figure.  We show the naive BFKL (lower curves)
and improved BFKL MC (upper curves) ratios for $p_T>50\; \GeV$ (dashed curves)
and 
$p_T>20\; \GeV$ (solid curves).  As at the Tevatron, we see that the
 BFKL MC predictions deviate strongly from those of naive BFKL due
 to nonleading effects.
The situation is slightly better than
at the Tevatron, where the  cross section was {never}  larger
at $1800\; \GeV$  than at $630\; \GeV$, in marked contrast to the
naive BFKL expectation.  Again, the LHC benefits from the larger
available phase space.

\section{Conclusions}

In summary, the ability of the LHC general purpose detectors
to measure `Mueller-Navelet' forward jets with large rapidity and
modest transverse momentum will allow an important test of QCD
`BFKL' physics. In this paper
we have presented the results of an improved BFKL Monte Carlo
calculation for 
dijet production at large rapidity separation at the LHC.  In this approach, 
subleading effects such as kinematic constraints and the running of the 
strong coupling constant $\as$ are incorporated by unfolding the implicit BFKL
sum over emitted gluons to make the sum explicit.\footnote{Another
potentially important effect, not included in the present study,
may come from the next-to-leading order perturbative corrections
to the BFKL equation, which have recently been calculated \cite{NLL}.}

We found that, as at the Tevatron \cite{os,osbis}, these subleading
effects can be substantial and in particular lead to some suppression
of gluon emission.  This kinematic suppression is however not as dramatic
at the LHC, because of its relatively higher center-of-mass energy compared
to the jet transverse momentum threshold.   As a result, improved BFKL MC
predictions tend to retain more BFKL-type behavior because of the 
greater phase space for emitting gluons.  We saw that as the $p_T$ threshold
was lowered, the BFKL MC results became more naive-BFKL-like.  This
was true for both the azimuthal decorrelation and the cross section ratio
at different collision energies.  

Noting that forward jets are also produced at the LHC in heavy Higgs boson 
production via gauge boson fusion, we compared the two dijet production
mechanisms.  The Higgs case is characterized by a lack of hadronic activity 
in the rapidity region between the two jets (the `rapidity gap'),
in contrast to the BFKL case where the region between the jets is
populated by emitted gluons which give rise to
an azimuthal decorrelation of the outside jets.  Interestingly, however,
we found that
the Higgs boson itself acts as a decorrelator and the jets produced 
in association with a heavy Higgs show a stronger decorrelation than
the BFKL jets.

The LHC holds promise for studies of dijet production in BFKL physics,
and such studies will benefit from jet detection capabilities that 
extend far into the forward region and allow for $p_T$ thresholds as low as
possible.  However, kinematic effects can still be of quantitative
importance, and should therefore be incorporated in the theoretical
analyses.

\vspace{1.0cm}
\noindent {\large \bf Acknowledgements} \\

\noindent 
This work was supported in part by the U.S. Department of Energy,
under grant DE-FG02-91ER40685 and by the U.S. National Science Foundation, 
under grant PHY-9600155.

\newpage


\end{document}